\newif\ifelsevier
\elsevierfalse

\ifelsevier
	\documentclass[preprint]{elsarticle}
\else
	\documentclass{acm_proc_article-sp}
\fi

\usepackage[pdfdisplaydoctitle=true,
colorlinks=true,
citecolor=blue,
linkcolor=blue,
urlcolor=blue]{hyperref}

\usepackage{textcomp}

\ifelsevier
	\journal{Future Generation Computer Systems}
\else
	\makeatletter
	\let\@copyrightspace\relax
	\makeatother
\fi

\begin{document}

\ifelsevier
	\begin{frontmatter}
\fi

\title{Marrying Many-core Accelerators and InfiniBand \\ for a New Commodity Processor}

\ifelsevier
\else
	\numberofauthors{2}
	\author{
	\alignauthor
	Konstantin S. Solnushkin\\
      		\email{konstantin@solnushkin.org}
	\alignauthor
	Yuichi Tsujita\titlenote{Present address: RIKEN, AICS, Japan}\\
      		\affaddr{Kinki University, Japan}\\
	       \email{yuichi\_tsujita@fw.ipsj.or.jp}
	}
	\maketitle
\fi

\begin{abstract}
During the last 15 years, the supercomputing industry has been using mass-produced, off-the-shelf components to build cluster computers. Such components are not perfect for HPC purposes, but are cheap due to effect of scale in their production. The coming exa-scale era changes the landscape: exa-scale computers will contain components in quantities large enough to justify their custom development and production.

We propose a new heterogeneous processor, equipped with a network controller and designed \textit{specifically} for HPC. We then show how it can be used for enterprise computing market, guaranteeing its widespread adoption and therefore low production costs.
\end{abstract}

\ifelsevier
	\begin{keyword}
	High-performance computing \sep Economics
	\end{keyword}
\else
	\category{C.1}{Computer Systems Organization}{Processor Architectures}
	\keywords{High-performance computing, Economics}
\fi

\ifelsevier
	\end{frontmatter}
\fi

\section{Introduction}
In 1990s, commodity off-the-shelf components allowed to build inexpensive but powerful cluster computers, disrupting the supercomputing market. Those components were not perfect for HPC, but were readily available and cheap. Current attempts to use commodity components are still focused on taking technologies initially created for enterprise computing and painfully fitting them into the Procrustean bed of HPC.

However, the future exa-scale computers will have so many identical building blocks such as CPUs -- on the order of millions \cite{dongarra2011international} -- that it becomes feasible to amortise their custom design and manufacturing costs over large production batches. Therefore we suggest to capitalise on this trend by designing a new commodity processor, with HPC being its primary workload.

At the same time, the enterprise computing market is significantly bigger than HPC market. Thus, for widest adoption, we should design the processor to make it suitable for data centre workloads as well, resulting in a unified architecture.

We believe that architecture of the new CPU should be based on the many-core paradigm, while getting rid of inefficiencies found in some of the current implementations. For example, the Intel Xeon Phi product received positive feedback from early adopters; however, the accelerator board is not ``standalone'', as it needs to be plugged into a ``host'' computer. This requires the use of a separate host CPU which leads to decreased density and loss of flexibility. Even more importantly, the accelerator communicates with the outside world through the host's network adaptor via PCIe connection, which adds to latency.

The current Xeon Phi implementation is not very useful for generic data centre applications, either. Configurations with up to 8 accelerator boards in a server were demonstrated, but in this case each accelerator receives only a share of network bandwidth.

We propose remedying the situation by integrating a many-core unit with a general-purpose multi-core CPU and a network adaptor, thus turning the board into a standalone and bootable server and cluster compute node. We then argue why this product can become useful not just to HPC but to a significantly broader user base, including data centre environments and desktop workstations. As predicted by HiPEAC \cite{duranton2010hipeac}, heterogeneous chips, similar to the one proposed in this paper, will be prevalent in the future.

\section{Related Work}

DAL Project (``Defying Amdahl's Law'') \cite{dal-project} explores future many-core architectures where several complex cores are accompanied by hundreds of simple cores on the same chip. The project's proposal is to intermittently clock complex cores at a high frequency, subject to thermal dissipation limit of the chip, to speed up execution of sequential parts of applications, while parallel parts continue running on the array of simple cores. The goal of the project is to improve microarchitecture so that the running thread can quickly migrate from a simple core to the complex one, and vice versa. As the heterogeneous CPU that we propose in this paper also features a mix of complex and simple cores, microarchitectural advances provided by the DAL project are applicable in our case.

DEEP Project (``Dynamical Exascale Entry Platform'') \cite{deep-project} proposes to build a ``cluster of accelerators'' -- that is, an array of independent, bootable accelerator boards connected with a high-speed network. This array is then connected to a conventional HPC cluster. Compute jobs are then diverted to the part of this complex which is most suitable for their execution. The structure of this ``cluster of accelerators'' is not intended for workloads other than HPC. Our approach is different: while we also propose to have independent and bootable accelerator boards, our CPU will contain cores complex enough to run nearly any workload, ranging from HPC to data mining and from CAE to data centre tasks.

``Project Denver'', proposed by NVIDIA \cite{nvidia-project-denver}, intends to build a heterogeneous processor that couples general-purpose cores based on ARM architecture with CUDA-programmed GPU cores. However, the GPU part is unable to run data centre server workloads. Another NVIDIA project is ``Echelon'' \cite{keckler2011gpus}, which its authors claim will become a general-purpose system suitable for data-intensive and HPC workloads and based on long instruction word (LIW) architecture.

``Runnemede'' chip, proposed by Intel \cite{carter2013runnemede}, employs a large number of simple cores called execution engines, controlled by a smaller number of general-purpose cores called control engines. Execution borrows ideas from dataflow architectures.

``Mont-Blanc Project'' \cite{rajovic2013low} investigates the use of commodity, readily-available ARM processors such as used in cell phones for HPC purposes.

\section{Structure and Functions}

\subsection{Integration Scenarios}

We describe three possible integration scenarios. In the simplest case, two mass-produced, off-the-shelf chips -- one for a multi-core CPU, and one for a many-core accelerator -- are placed on a single board, accompanied by a network chip and memory chips. The board is bootable and serves as a standalone cluster compute node. This is a step forward from the current Intel Xeon Phi implementation that needs a host computer to be plugged into, for booting and communicating with the outside world.

However, this scenario lacks efficient communication between the multi-core and many-core parts. Hence the second, more advanced integration scenario is to put multi-core and many-core parts into a single IC package, or possibly on the same die, with a network adapter still implemented as a separate chip. Fast communication between two parts of a tandem via an on-chip network will allow the multi-core part to perform I/O and MPI delegation functions for the many-core.

Finally, the third scenario integrates the network adaptor on the same chip. Deep integration, resulting in a System-on-a-Chip (SoC), will incur custom design and fabrication costs which are best offset by mass production. However, we believe that applicability of this SoC for a very broad market will help amortise costs and turn it into a new sort of inexpensive commodity hardware.

To make the SoC suitable for desktop workstations as well, we propose to add a simple GPU and some commonly used accelerators, such as hardware-assisted encryption units; this will not take much of the die space. These units may remain unused for some workloads (being so-called ``dark silicon''; similar to how floating-points units of current CPUs are mostly unused in data-mining and server workloads \cite{taylor2012dark}). Adding an FPGA unit on the chip, similar to the ``Xilinx Zynq'' product, will open yet more field-programming flexibility.

\begin{figure*}
\centering
\ifelsevier
	\includegraphics[width=120mm]{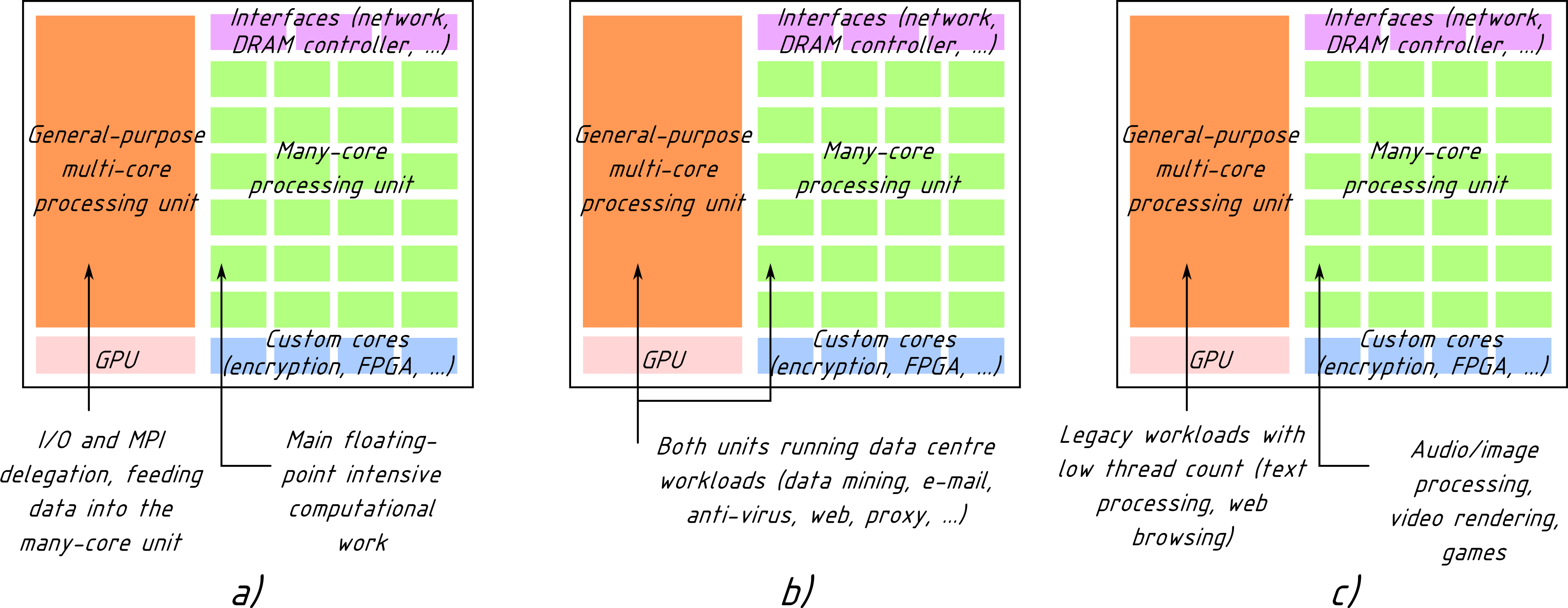}
\else
	\includegraphics[width=150mm]{fig_01}
\fi
\caption{Three use cases for the proposed system with corresponding workloads: (a) HPC environments, (b) generic data centre environments, (c) desktop computing environments}
\label{fig:use-cases}
\end{figure*}

\subsection{Use Cases}

The many-core part will consist of identical multithreaded simple low-power cores tailored to floating-point computations, like in existing many-core implementations such as Intel Xeon Phi and GPUs.

Three use cases for utilising the proposed chip in different environments are presented on Figure \ref{fig:use-cases}. In each of these scenarios, the general-purpose multi-core unit doesn't have to be very powerful in terms of floating-point performance, as it is assumed that for intensive computing tasks the software will be able to utilise the many-core part (via, say, OpenCL).

For data centre workloads, cores in the many-core part can be visible to the operating system as individual CPUs for easy task scheduling. This usage, involving multithreaded simple cores, is similar to Sun's ``UltraSPARC T2'' CPU and a recent Hewlett-Packard's ``Project Moonshot'' \cite{hp-project-moonshot-theregister}.

To reduce requirements on memory bandwidth, cache memory will be required for legacy applications; we propose to implement it using eDRAM technology. The effectiveness of this approach was proven by IBM's POWER7 processor \cite{kalla2010power7}. Alternatively, for newer applications that prefer to manage memory accesses on their own, and don't require hardware caching logic, this on-chip memory can be configured as software-controlled scratch pads, leading to energy savings \cite{keckler2011gpus, carter2013runnemede}.

The crucial question is how many cores the many-core part should have. Chip's I/O bandwidth is limited by its pin count, and we should aim for ample bandwidth per core, as the chip's primary purpose is HPC. Therefore we shouldn't be tempted to place as many cores as possible. Other limits on the number of cores are related to heat dissipation and yield of the semiconductor fabrication process.

Our choice of network is InfiniBand, which currently can be recognised as commodity technology. It has the following benefits: reasonable technical characteristics, clear technology roadmap, hardware is produced on a large scale, and it supports a variety of network topologies. For HPC and data centre environments, each 18 boards with the proposed CPU can be connected, via a backplane, to the InfiniBand switch chip, which would further connect them to the rest of the fabric (in fat-tree, torus or other topologies).

Besides the aforementioned components, the motherboard will contain DRAM memory modules as well as an optional flash-based module for scratch storage, creating an additional level of memory hierarchy, as was featured in ``Gordon'' supercomputer \cite{strande2012gordon}.

When running at full speed, the proposed chip is best cooled with water. The feasibility and reliability of this approach on the large scale was verified with the ``SuperMUC'' machine \cite{brehm2012energy}. To facilitate waste heat reuse in water-cooled environments, mid-scale installations ($\sim$100 kW and higher) are preferred \cite{solnushkin-fruits-of-computing}. However, in desktop computing water cooling is usually not available, and aggressive automatic power throttling (via under-clocking) of the many-core part will be required during its operation to prevent accidental chip overheating (``dim silicon'' approach, according to Taylor's classification \cite{taylor2012dark}).

\section{Software Ecosystem}

Instead of the ubiquitous but outdated and proprietary X86 architecture, it is tempting to utilise open-source architecture in the proposed system. This will promote research and collaboration, including parties from the private sector, and can steer competition.

A good candidate is the ``OpenSPARC'' architecture and the ``OpenSPARC T2'' microarchitecture implementation. (For example, ``SPARC64 VIIIfx'' CPU, as found in the Japanese ``K Computer'', is also based on SPARC architecture \cite{yoshida2012sparc64}). To make a leap to the many-core, the same trick could be employed that Intel used for Xeon Phi: stripping less-needed features such as out-of-order execution while simultaneously widening floating-point units.

The OpenSPARC architecture is supported by GNU/Linux operating system and the GCC (GNU Compiler Collection). In the HPC environment, the amount of efforts required to build the ecosystem for the new CPU (upgrading the Linux kernel, porting MPI implementations and several important numerical libraries and advanced compilers) is lower compared to the server segment, where additional usual data centre applications will need to be ported for market acceptance. Proliferation in the desktop segment is difficult to achieve unless Microsoft Windows and device drivers are ported to OpenSPARC. Instead, a GNU/Linux distribution such as Ubuntu running on OpenSPARC could be refined to a notable point. 

Using OpenSPARC as the base architecture may not lead to designs achieving as low performance per watt as Intel Xeon Phi or IBM BlueGene/Q designs, although heat reuse methods will alleviate this problem. What is more important, however, is that the proposed CPU will find use in many niches; therefore its mass production will keep its price low. Additionally, methods and technologies from the HPC niche will be directly applicable in the data centre field due to unified architecture.

There is a risk that a product that tries to fit all three niches -- HPC, server and desktop -- might fit in none. However, the market is diluted by assorted products that appear every year. Perhaps it's time for reunification. Adherence to open-source policies at all stages -- in both accepting and giving -- will help offset the costs and reach the goal faster.

\begin{figure}
\centering
\includegraphics[width=80mm]{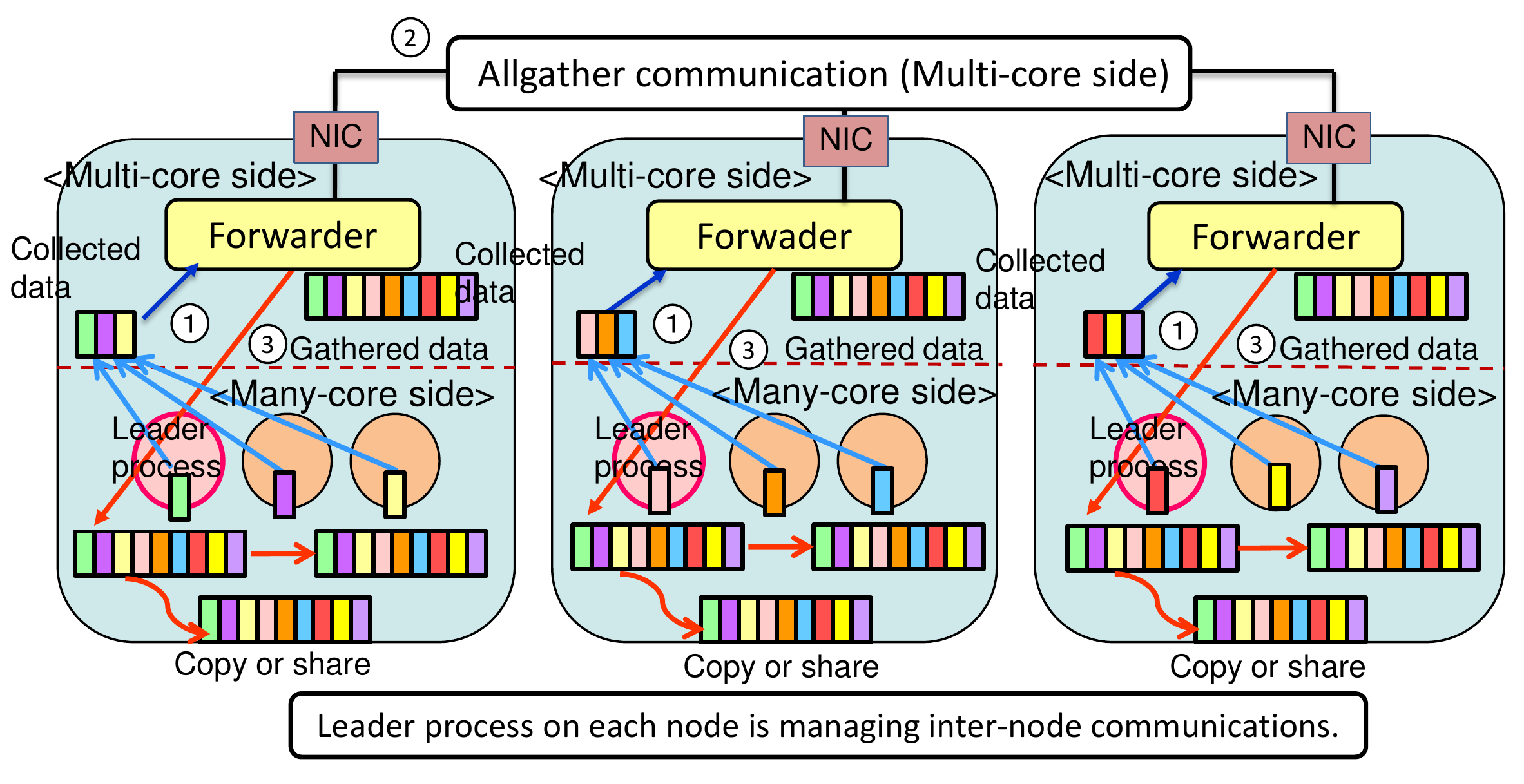}
\caption{Optimisation scheme in collective communications, in a hybrid system}
\label{fig:allgather-delegation}
\end{figure}

\begin{figure}
\centering
\includegraphics[width=80mm]{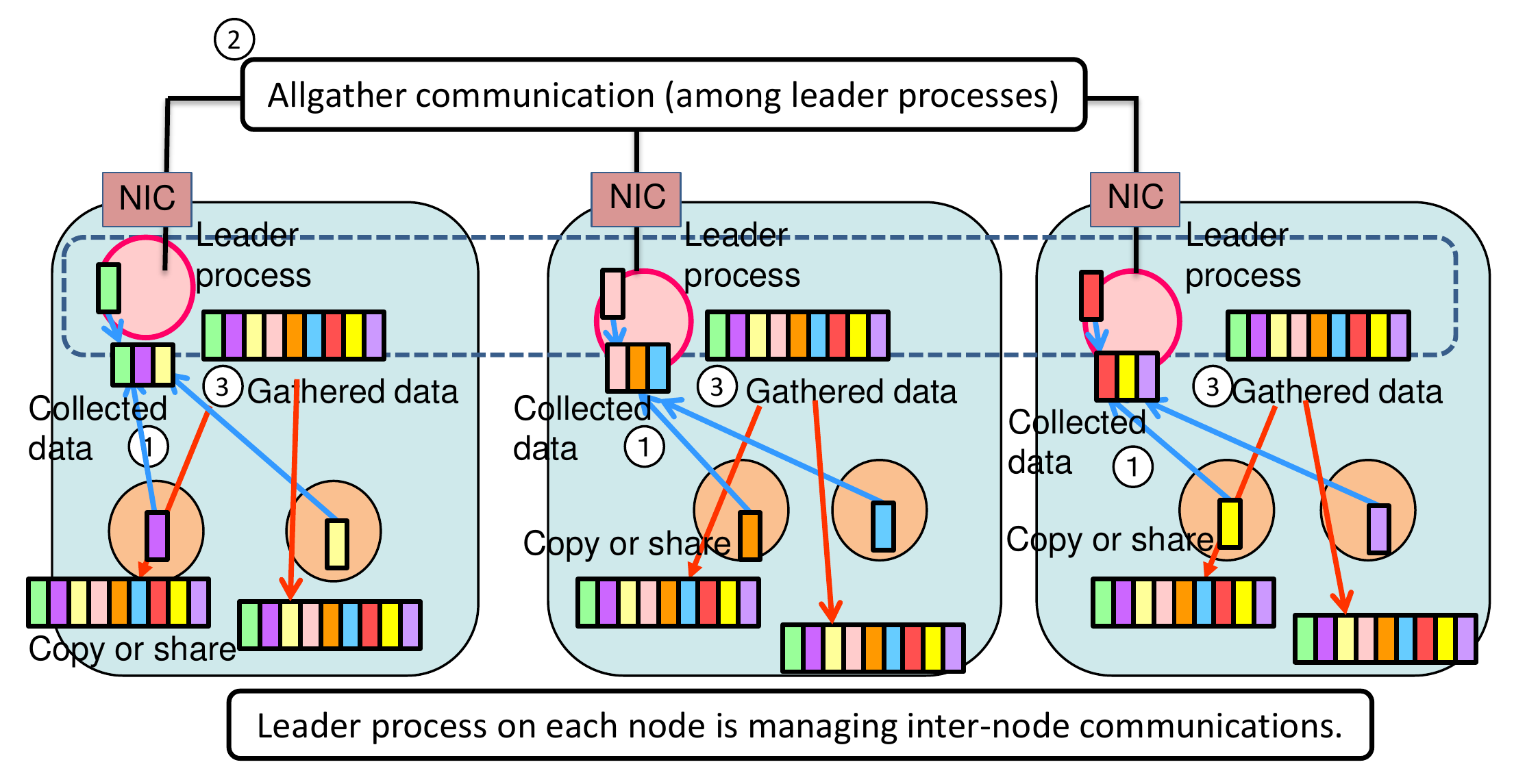}
\caption{Optimisation scheme in collective communications, many-core processors only}
\label{fig:allgather-bootable}
\end{figure}

\section{Perspective View of Collective Operation Implementations}
Flat MPI is no longer scalable on recent accelerator-based hybrid systems where many-core processors are provided in the form of a PCIe card \cite{yoshinaga2012delegation}. Thus we propose a hybrid system, with many-core and multi-core processors within the same compute node. However, MPI is still useful for most applications.

Therefore we describe here a perspective view of the MPI library software stack, especially for collective operations. Existing software tools for designing computer clusters \cite{solnushkin2012computer} can be enhanced to accompany their designs with hints on optimal MPI process placement among cluster nodes, taking network proximity into account. This will facilitate efficient operation of multi-layer MPI communicators and collective operations.

Providing an MPI library for such a hybrid system may attract attention of existing application users, allowing them to easily exploit parallelism of the hybrid systems. Here we show the following schemes, with MPI\_Allgather operation as an example.

\begin{enumerate}
\item A delegation mechanism as shown in Figure \ref{fig:allgather-delegation}
\begin{itemize}
\item Considering the recent hybrid architecture with many-core and multi-core processors within the same node
\item  Aggregating collective communication and I/O requests from the many-core side, followed by real operations by a forwarder process on the multi-core side
\item Gathered data are then copied to non-leader processes, or accessed by them in a shared manner.
\end{itemize}

\item Multi-layered system available on many-core only system (multi-core unit not available), shown in Figure~\ref{fig:allgather-bootable}. Here we have the following assumptions about this system:
\begin{itemize}
\item Complicated hierarchy in memory architecture
\item Shortage of available memory per core\\
Therefore a multi-layered MPI communicator management is beneficial. Based on this scheme, we may form groups within compute nodes:
\item Every group has internal collective communications.
\item Every leader process manages communications on behalf of the associated group.
\item Gathered data are then copied to non-leader processes, or accessed by them in a shared manner.
\end{itemize}
\end{enumerate}

In both cases, some kind of shared memory management mechanism is required inside a compute node.

\begin{figure}
\centering
\ifelsevier
	\includegraphics[width=120mm]{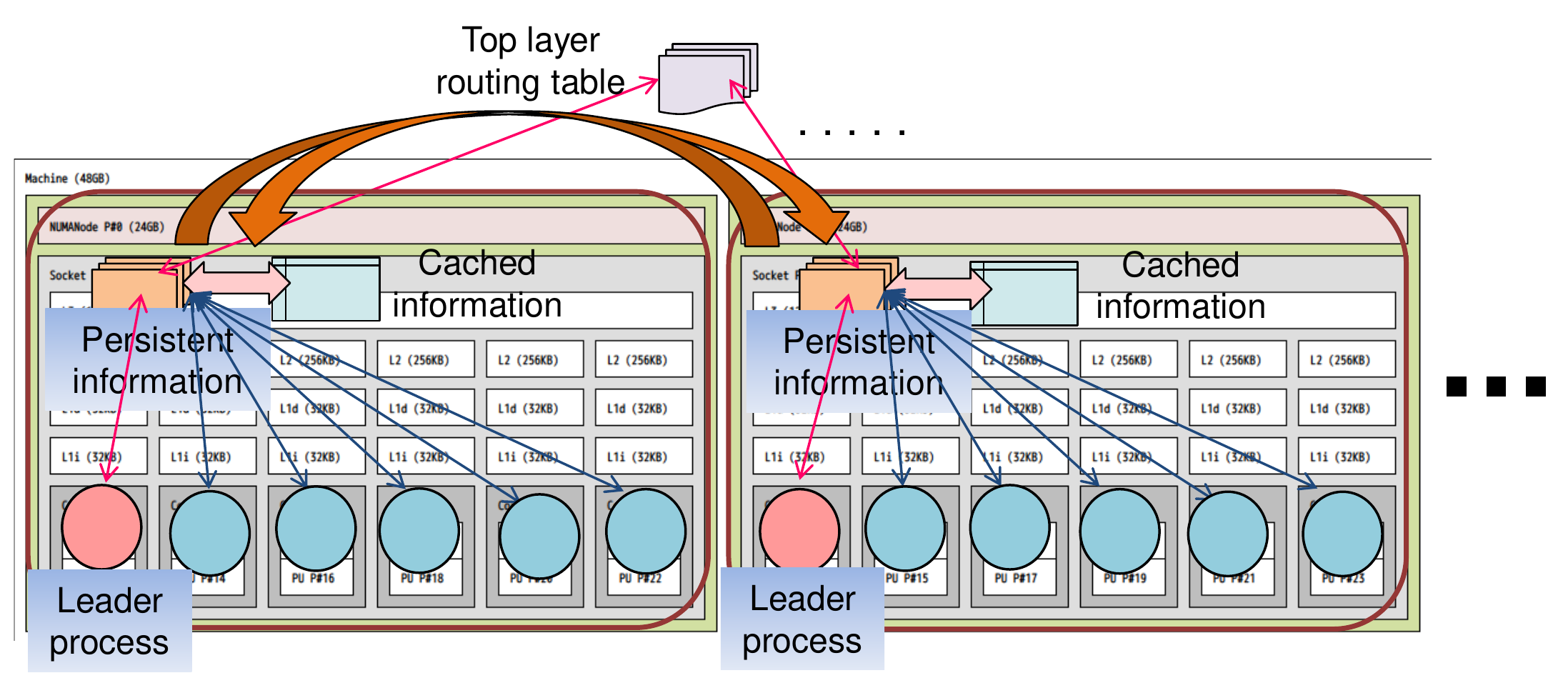}
\else
	\includegraphics[width=80mm]{loc-aware-communicator}
\fi
\caption{Communication optimisation based on hardware affinity}
\label{fig:loc-aware-communicator}
\end{figure}

Another optimisation will come from making use of hardware affinity in process placement and communication topology. ``hwloc'' \cite{broquedis2010hwloc} or ``likwid'' \cite{treibig2010likwid} are the candidates for understanding the memory hierarchy. These or similar tools can be used to facilitate effective process mapping or distributed MPI communicator information management, as shown in Figure \ref{fig:loc-aware-communicator}.

APIs of such hardware affinity tools can be used in process management scheme of an extended MPI library. In order to minimise local communicator information on each node, only information about processes in the local group is kept inside the same node. Information of external processes is initially queried through communicator management scheme, and cache mechanism will be implemented to reuse connection information.

Combining interconnection fabrics inside the same chip may lead to smaller communication latency, because we can eliminate PCIe access overhead. However, this approach has technical challenges associated with custom fabrication. Having many functions in the same chip leads to a higher electric power consumption of the chip; this is yet another challenge.

\section{Conclusions}
In this paper we propose a new commodity processor, designed with HPC in mind but suitable for a much broader set of workloads. We review a perspective approach to the MPI library software stack, where a multi-core unit performs collective operations on behalf of the many-core unit. We also note that the appeal of the resulting hardware solution to the data centre market will justify its large scale production, keeping costs low.

\section{Acknowledgements}
The work was \href{http://www.iccs-meeting.org/iccs2013/papers/schedule.php?session=W63c}{presented} at the International Conference on Computational Science (ICCS 2013) in Barcelona, Spain. Travel support for Konstantin S. Solnushkin was provided by the National Research University ITMO, St. Petersburg, Russia under agreement \textnumero 11.G34.31.0019.


\ifelsevier
	\bibliographystyle{elsarticle-num}
\else
	\bibliographystyle{abbrv}
\fi

\bibliography{../text/text/bibliography/literature}

\begin{thebibliography}{10}
\expandafter\ifx\csname url\endcsname\relax
  \def\url#1{\texttt{#1}}\fi
\expandafter\ifx\csname urlprefix\endcsname\relax\def\urlprefix{URL }\fi
\expandafter\ifx\csname href\endcsname\relax
  \def\href#1#2{#2} \def\path#1{#1}\fi

\bibitem{dongarra2011international}
J.~Dongarra, P.~Beckman, T.~Moore, P.~Aerts, G.~Aloisio, J.~Andre, D.~Barkai,
  J.~Berthou, T.~Boku, B.~Braunschweig, et~al., The international exascale
  software project roadmap, International Journal of High Performance Computing
  Applications 25~(1) (2011) 3.

\bibitem{duranton2010hipeac}
M.~Duranton, S.~Yehia, B.~De~Sutter, K.~De~Bosschere, A.~Cohen, B.~Falsafi,
  G.~Gaydadjiev, M.~Katevenis, J.~Maebe, H.~Munk, N.~Navarro, A.~Ramirez,
  O.~Temam, M.~Valero, The {HiPEAC} vision, \url{http://www.hipeac.net/roadmap}
  (2010).

\bibitem{dal-project}
{DAL Project}, Defying {A}mdahl's law, \url{http://www.irisa.fr/alf/dal/}.

\bibitem{deep-project}
{DEEP Project}, Dynamic exascale entry platform,
  \url{http://www.deep-project.eu/}.

\bibitem{nvidia-project-denver}
B.~Dally, ``{P}roject {D}enver'' processor to usher in new era of computing,
  \url{http://blogs.nvidia.com/2011/01/project-denver-processor-to-usher-in-new-era-of-computing/}
  (January 2011).

\bibitem{keckler2011gpus}
S.~W. Keckler, W.~J. Dally, B.~Khailany, M.~Garland, D.~Glasco, {GPU}s and the
  future of parallel computing, Micro, IEEE 31~(5) (2011) 7--17.

\bibitem{carter2013runnemede}
N.~P. Carter, A.~Agrawal, S.~Borkar, R.~Cledat, H.~David, D.~Dunning,
  J.~Fryman, I.~Ganev, R.~A. Golliver, R.~Knauerhase, R.~Lethin, B.~Meister,
  A.~K. Mishra, W.~R. Pinfold, J.~Teller, J.~Torrellas, N.~Vasilache,
  G.~Venkatesh, J.~Xu, Runnemede: An architecture for ubiquitous
  high-performance computing, in: High Performance Computer Architecture
  (HPCA), 2013 IEEE 19th International Symposium on, 2013.

\bibitem{rajovic2013low}
N.~Rajovic, L.~Vilanova, C.~Villavieja, N.~Puzovic, A.~Ramirez, The low power
  architecture approach towards exascale computing, Journal of Computational
  Science\href {http://dx.doi.org/10.1016/j.jocs.2013.01.002}
  {\path{doi:10.1016/j.jocs.2013.01.002}}.

\bibitem{taylor2012dark}
M.~B. Taylor, Is dark silicon useful? {H}arnessing the four horsemen of the
  coming dark silicon apocalypse, in: Proceedings of the 49th Annual Design
  Automation Conference, ACM, 2012, pp. 1131--1136.

\bibitem{hp-project-moonshot-theregister}
T.~P. Morgan, {HP} {P}roject {M}oonshot hurls {ARM} servers into the heavens,
  \url{http://www.theregister.co.uk/2011/11/01/hp_redstone_calxeda_servers/}
  (November 2011).

\bibitem{kalla2010power7}
R.~Kalla, B.~Sinharoy, W.~J. Starke, M.~Floyd, {POWER7}: {IBM}'s
  next-generation server processor, IEEE Micro 30~(2) (2010) 7--15.

\bibitem{strande2012gordon}
S.~Strande, P.~Cicotti, R.~Sinkovits, W.~Young, R.~Wagner, M.~Tatineni,
  E.~Hocks, A.~Snavely, M.~Norman, Gordon: design, performance, and experiences
  deploying and supporting a data intensive supercomputer, in: Proceedings of
  the 1st Conference of the Extreme Science and Engineering Discovery
  Environment: Bridging from the eXtreme to the campus and beyond, ACM, 2012,
  p.~3.

\bibitem{brehm2012energy}
M.~Brehm, A.~Auweter, H.~Huber, T.~Wilde, Energy efficient {HPC} systems:
  Concepts, procurement \& installation, in: Proceedings of International
  Supercomputing Conference, ISC'12, 2012.

\bibitem{solnushkin-fruits-of-computing}
K.~S. Solnushkin, Fruits of computing: Redefining '{G}reen' in {HPC} energy
  usage,
  \url{http://clusterdesign.org/2012/08/fruits-of-computing-redefining-green-in-hpc-energy-usage/}
  (August 2012).

\bibitem{yoshida2012sparc64}
T.~Yoshida, M.~Hondo, R.~Kan, G.~Sugizaki, {SPARC64} {VIIIfx}: {CPU} for the
  {K} computer, Fujitsu Sci. Tech. J 48~(3) (2012) 274--279.

\bibitem{yoshinaga2012delegation}
K.~Yoshinaga, Y.~Tsujita, A.~Hori, M.~Sato, M.~Namiki, Y.~Ishikawa,
  Delegation-based {MPI} communications for a hybrid parallel computer with
  many-core architecture, in: Recent Advances in the Message Passing Interface,
  LNCS 7490, Springer, 2012, pp. 47--56.

\bibitem{solnushkin2012computer}
K.~S. Solnushkin, \href{http://konstantin.solnushkin.org}{Computer cluster
  design automation using web services}, in: Proceedings of International
  Supercomputing Conference, ISC'12, 2012.
\newline\urlprefix\url{http://konstantin.solnushkin.org}

\bibitem{broquedis2010hwloc}
F.~Broquedis, J.~Clet-Ortega, S.~Moreaud, N.~Furmento, B.~Goglin, G.~Mercier,
  S.~Thibault, R.~Namyst, hwloc: A generic framework for managing hardware
  affinities in {HPC} applications, in: Parallel, Distributed and Network-Based
  Processing (PDP), 2010 18th Euromicro International Conference on, IEEE,
  2010, pp. 180--186.

\bibitem{treibig2010likwid}
J.~Treibig, G.~Hager, G.~Wellein, Likwid: A lightweight performance-oriented
  tool suite for x86 multicore environments, in: Parallel Processing Workshops
  (ICPPW), 2010 39th International Conference on, IEEE, 2010, pp. 207--216.

\end{thebibliography}

\end{document}